# A NESTED GENETIC ALGORITHM STRATEGY FOR THE OPTIMAL PLASTIC DESIGN OF FRAMES

A. Greco[a], F. Cannizzaro[a], R.Bruno[b], A. Pluchino[b,c]

[a]*Department of Civil Engineering and Architecture,*
*University of Catania, viale A. Doria 6, Catania, Italy*
[b]*Sezione INFN of Catania, via S.Sofia 64, Catania, Italy*
[c]*Department of Physics and Astronomy "E.Majorana" University of Catania*

email: *annalisa.greco@unict.it,*
*francesco.cannizzaro@unict.it, riccardo.bruno@ct.infn.it, alessandro.pluchino@ct.infn.it*

## Abstract

An innovative strategy for the optimal design of planar frames able to resist to seismic excitations is here proposed. The procedure is based on genetic algorithms (GA) which are performed according to a nested structure suitable to be implemented in parallel computing on several devices. In particular, this solution foresees two nested genetic algorithms. The first one, named 'External GA', seeks, among a predefined list of profiles, the size of the structural elements of the frame which correspond to the most performing solution associated to the highest value of an appropriate fitness function. The latter function takes into account, among other considerations, of the seismic safety factor and the failure mode which are calculated by means of the second algorithm, named 'Internal GA'. The details of the proposed procedure are provided and applications to the seismic design of two frames of different size are described.

**Corresponding author**: A.Greco, e-mail: annalisa.greco@unict.it, tel: +390957382251

# 1  Introduction

The design of structures is a widely debated topic in the last years, due to its complexity, to the evolution of the analysis methods and to the conceptual advances made in the last decades towards a deeper understanding of the structural behaviour of buildings.

The design of a structure may be faced considering different aspects and strategies with increasing levels of complexity and elaborateness. Usually, for new constructions, the codes [1] suggest the employment of global elastic analysis and then to singularly design each member. However, a local design does not guarantee to have an effective global behaviour of the building, especially in seismic conditions. For the latter reason, modern strategies of design of structures must be respectful of the concepts of global ductility which ensures that, during an earthquake, a structure is able to dissipate an adequate amount of energy before collapsing. Aiming at favouring a ductile behaviour of a structure, some tools, such as the structural factor to reduce the design spectrum and the regulations associated to the strength hierarchy [2] of beams and columns, have been more recently introduced in the design process.

Usually, especially in the case of steel structures, the economic criterion (i.e. minimum weight) is considered as the only variable to be minimized in the design process, as long as the structural requirements are satisfied. Within this context, the design problem consists in finding a set of variables, namely the cross section of the members, which minimizes a certain function (weight of the structure) under some design constraints, which have to be evaluated following a structural analysis. The latter aspect usually prevents the formulation of explicit approaches for a comprehensive design of the structural members of buildings. Under these hypotheses, and considering the adoption of linear analyses, the employment of evolutionary algorithms [3] turned out to be very useful to guide the design process [4-7].

However, the adoption of a nonlinear approach for the design of structures, even in presence of new constructions, has been suggested by several authors both considering static [8,9] and dynamic [10,11] load conditions. The adoption of nonlinear analyses was also combined in [12] with a multi-objective approach, able to account not only for the minimization of the weight but also for other performance objectives. Nonlinear analyses guarantee a deeper knowledge of structures and are preferred in the study of existing structures; however, the high required computational effort and the number of configurations that are usually needed to be analysed in a design process, make them usually unsuitable for a global speedy design of structures.

A fair compromise, in terms computational effort, between linear and nonlinear (pushover and time-history) analyses, is represented by the limit analysis. Plastic design of structures has been introduced since the early 1950 [13-17] and never stopped developing. Limit analysis is able to account, in a design process, for the actual global resistance of a structure subjected to a certain load condition with a reasonable computational cost. The plastic design of structures allows not only to take into account the global resistance of the structure, but also provides useful information in terms of failure mode [18,19]. The latter aspect was taken into account in [20] where a failure mode control was introduced in the design process as well as the second order effects and the hierarchy strength principle. Comprehensive overviews of the performance-based plastic design methods and, more in general, for the design of steel structures can be found in [21] and [22], respectively. However, in the plastic design of a structure it is difficult to assure its desired global performance when a ductility demand has to be guaranteed.

In the field of the limit analysis, some of the authors recently proposed a strategy for the evaluation of the ultimate load of frame structures based on the application of evolutionary algorithms to the method of the combination of elementary mechanism [23]. The proposed methodology can be applied both to regular [24,25] and irregular [26] frames. A further advance was then proposed in [27], where the second order effects [28] were introduced and a simplified capacity curve of a frame is introduced to evaluate its seismic capacity. The procedure illustrated in [27] is able not only to provide information on the resistance and failure mode of a frame, but also to assess in a simplified manner its displacement capacity, by evaluating the ultimate displacement of the control point according to the achievement of the limit chord rotation at each of the plastic hinges involved in the collapse mechanism.

In the present paper the procedure proposed in [27], which showed to be reliable and fast and is based on a genetic algorithm (GA), represents the internal routine (called "Internal GA") of a nested genetic algorithm strategy, able to design a frame with given geometry and loads. The external algorithm (called "External GA") selects the cross sections associated to the members of the frame, and evolves towards the configuration which shows the highest fitness. The performance of each of the selected frame configurations is evaluated not only according to an economic factor (minimum weight), but also considering other meaningful aspects, that is the safety factor, a hierarchy criterion and privileging those frames associated to a more dissipative failure mode. The strategy is implemented through a dedicated software developed within the agent-based multiplatform environment Netlogo [29] and it has been integrated into a Cloud based computational workflow built on top of the GARR Cloud Infrastructure, using modern and innovative strategies allowing to easily reproduce this work in accordance to the FAIR principles in OpenScience.

The conceived approach is a multi-objective design strategy, respectful of the main modern concepts for an effective design of structures, but also ensuring a significant computational advantage with respect to other nonlinear approaches based on pushover and nonlinear time history analyses. In addition, it is keen to introduce additional design criteria and to vary the relative weights of the adopted criteria. The procedure is validated against two significant case studies, which are duly analysed considering the needed computational effort, and demonstrating how the designed frames are those which maximize the defined fitness.

## 2 The proposed design strategy

In the present study planar regular frames with columns clamped at the base are considered. These are characterized by the number of floors $N_f$ and the number of columns $N_c$. The length of all the beams is $L$ and the height of the columns is $H$. A constant cross section for the beams within the same floor is considered and the same assumption is made for the columns. The frame may be loaded, at each floor, by concentrated horizontal forces $F_k$ and vertical distributed loads $q_k$.

With the aim of simulating seismic inputs, the horizontal forces are considered variable while the vertical loads are assumed to be distributed and of constant value within the same floor (Figure 1).

In the proposed design procedure, the geometry of the frame in terms of interstorey height and frame bays, is assigned. In addition, the vertical loads, and the horizontal ones which depend on the weights, are considered known as well. The unknowns are represented by the cross sections associated to the members, that is their stiffness and resisting plastic moments.

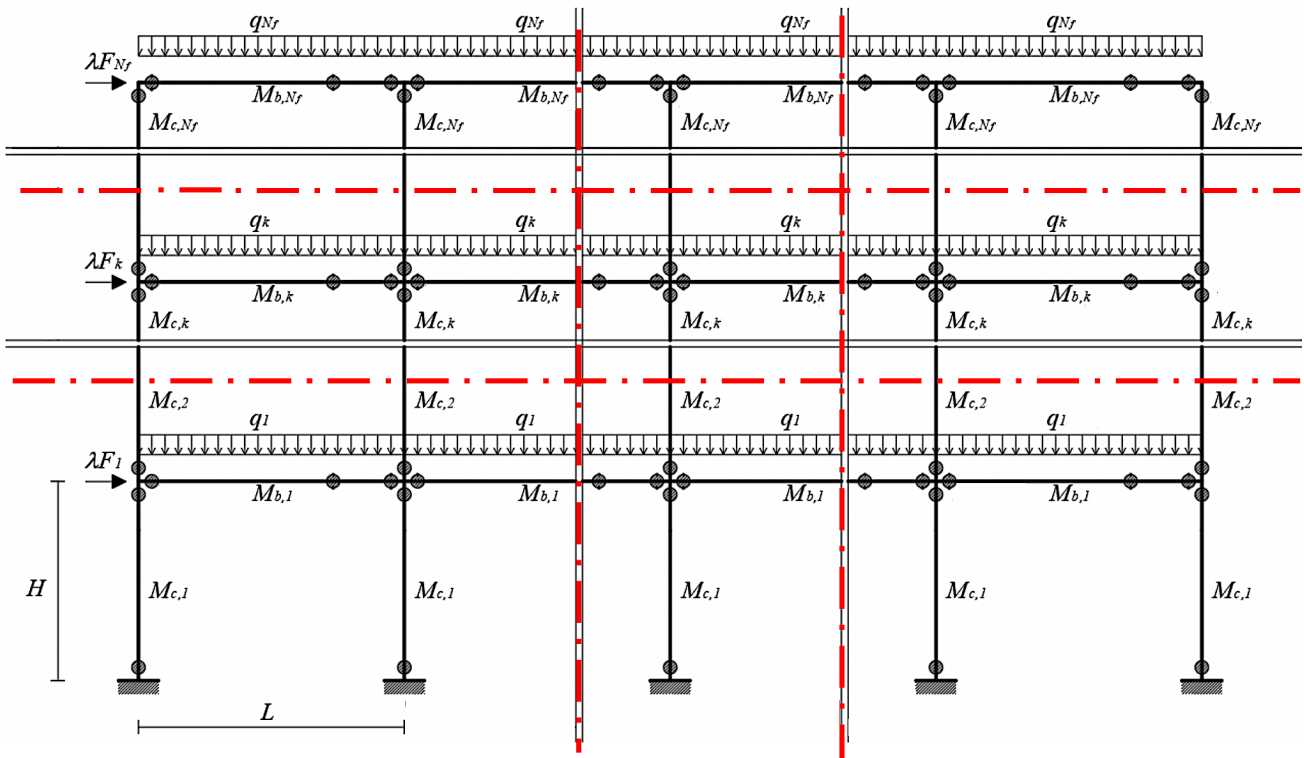

Figure 1. Layout of a generic planar frame

The proposed design approach, consisting in the two nested genetic algorithms, is described in the following subsections. Specifically, in the first subsection, the External GA, devoted to the identification of the best set of cross sections to be assigned to the members of the frame, is described. In the second subsection, the seismic assessment procedure proposed in [27], which constitutes the Internal GA, is recalled. In the third subsection, the procedure for the definition and calculation of the fitness of the generic frame, and the considerations which lead to the identification of the optimal design, are illustrated. The whole procedure is summarized in the flow chart reported in Figure 2.

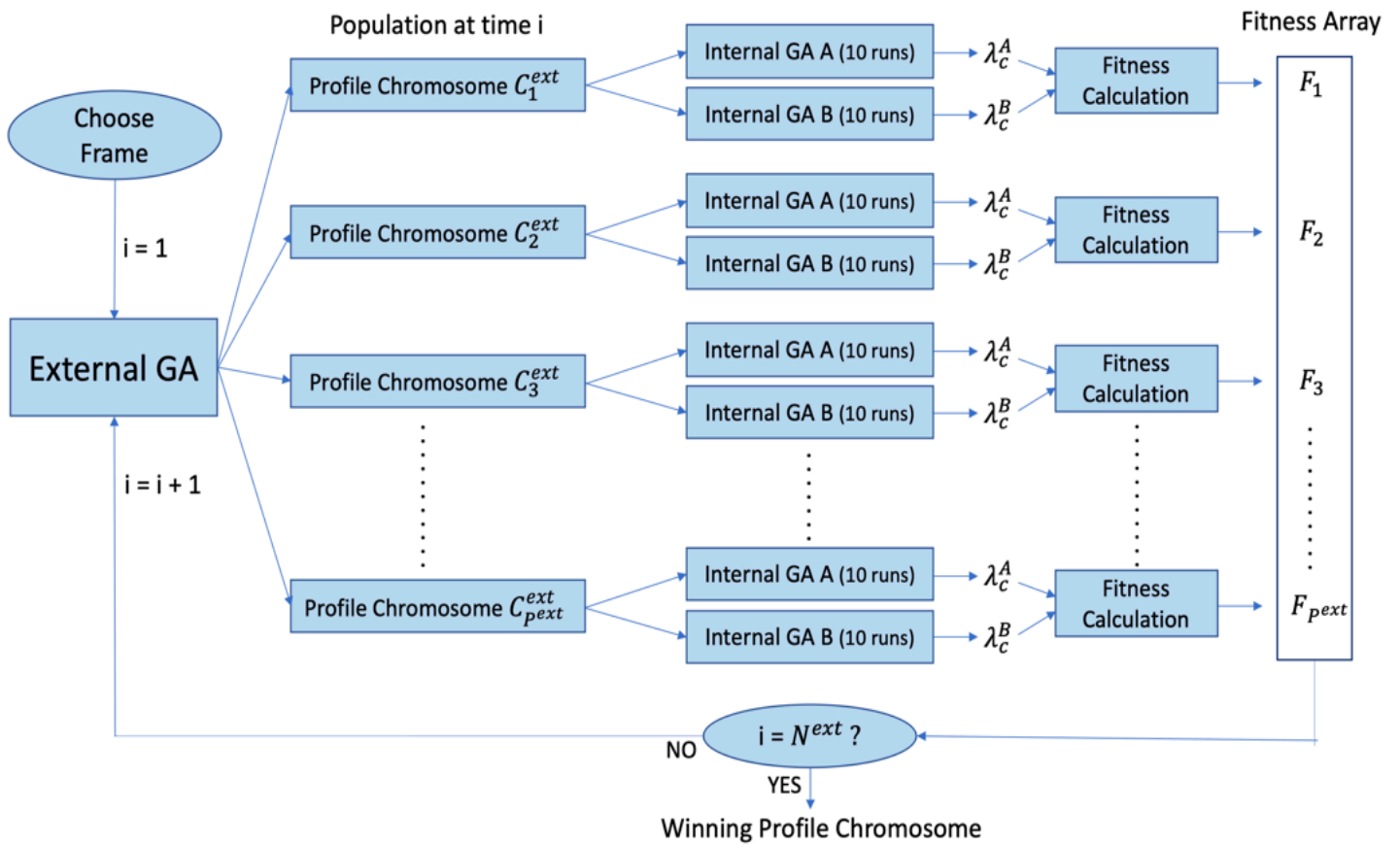

Figure 2. Flow chart of the nested algorithm.

## 2.1 The External GA (Design algorithm)

The main loop of the general procedure, here addressed as external, is devoted to the definition of the cross sections to be assigned to the members of the frame. The cross sections of the structural members are assumed to be variable in a predefined list of commercial profiles for steel frames or, in the case of reinforced concrete structures, in a predefined list of cross sections conveniently chosen. Since the procedure adopted in the present paper, in order to identify the optimal set of profiles, makes use of genetic algorithms, a population of profile chromosomes is considered. Each individual of the population is coded as a string of integer numbers, where each number (called "gene") represents the considered commercial profile within the predefined list. Therefore, a generic profile chromosome $C_k^{ext}$ of the population ($k = 1, ..., P^{ext}$) can be coded in the following string made of pairs of genes:

$$C_k^{ext} \equiv (c_{1c}, c_{1b}, c_{2c}, c_{2b}, ... c_{ic}, c_{ib}, ..., c_{Nfc}, c_{Nfb}) \qquad (1)$$

where $c_{ic}$ and $c_{ib}$ in each pair represent the identification codes of the cross sections of the columns and of the beams of the *i*-th floor, respectively. Each chromosome in the considered population has therefore $2N_f$ genes.

For each profile chromosome, and therefore for each generic frame, the correspondent seismic performance is evaluated by means of a limit analysis strategy procedure described in detail in [27], which represents the internal algorithm in the proposed nested strategy and is briefly recalled in the next subsection 2.2. The seismic performance of the frame, together with other engineering requirements, contribute to the definition of the fitness of the profile chromosome and therefore to the identification of the optimal set of cross sections, as described in subsection 2.3.

## 2.2 The Internal GA (Seismic assessment algorithm)

Given a profile chromosome $C_k^{ext}$, the load factor $\lambda_o$ of the horizontal forces is calculated following an approach which takes into account global collapse mechanisms obtained by means of linear combinations of three elementary ones: floor, beam and node mechanisms [15].

For each collapse mechanism, making use of the virtual work theorem, the value of the load multiplier $\lambda_o$ is given by:

$$\lambda_0 = \frac{W_{int} - W_{extV}}{W_{extH}} \qquad (2)$$

where $W_{extH}$ and $W_{extV}$ represent the work done by the horizontal forces and the vertical permanent load, respectively, and $W_{int}$ is the internal work concentrated in the plastic hinges. Analysing all the possible combinations of $N$ elementary mechanisms, the minimum value of $\lambda_o$ must be sought in order to obtain the real collapse load.

The minimization procedure makes again use of genetic algorithms through the definition of the population of "internal chromosomes" $C_h^{int}$ ($h = 1, ..., P^{int}$) defined as:

$$C_h^{int} \equiv (c_1, c_2, c_3, ..., c_j, ..., c_N) \qquad (3)$$

where $c_j = 1$ or $c_j = 0$ depending on fact that the $j$-th elementary mechanism is involved in the combination or not. For these internal algorithms, a number $N^{int}$ of generations will be considered. Once the winning internal chromosome is sought through the procedure proposed in [24], the corresponding final collapse load $\lambda_c$ can be obtained by decreasing the load factor $\lambda_o$ taking into account the second order effects of the work done by the vertical loads as follows:

$$\lambda_c = \lambda_0 - \gamma\delta \qquad (4)$$

where $\delta$ is the horizontal displacement of the top of the frame and $\gamma$ is the slope of the linear post-elastic descending branch accounting for the second order effects [27].

An approximated bilinear capacity curve, referred to a monitored point conveniently chosen at the top floor of the frame is adopted. In Figure 3 the construction of the proposed bilinear capacity curve (black line), which approximates the actual nonlinear one (red line) is depicted.

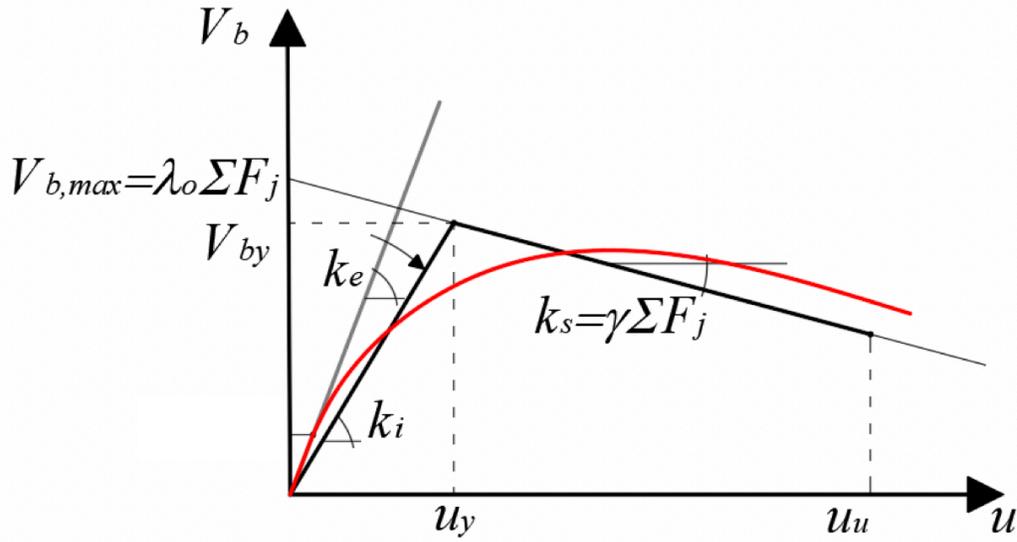

Figure 3. Approximated bilinear capacity curve [27]

It is worth higligting that, in order to take into account the progressive damage diffusion occurring in the frame during the collapse, the initial stiffness $k_i$ is an appropriate reduction of the elastic stiffness $k_e$ as described in [27]. The post-peak slope $k_s$ is evaluated considering the second order effects and is equal to $k_s = -\gamma \sum_{j=1}^{N_f} F_j$.

The capacity curve is then truncated considering as ultimate displacement $u_u$ the lowest among those associated to the achieving of limit states corresponding to two several commonly accepted principles, namely the global base shear reduction and the achieving of ultimate chord rotations in beams.

The described approximated bilinear curve is representative of the MDOF system and therefore of the whole frame. According to the procedure reported in [27], the properties of an equivalent SDOF system and the demand and capacity associated to the Near Collapse (NC) Limit State are compared in order to evaluate the safety factor. The latter is defined as the ratio between the ultimate displacements of the equivalent elastic-plastic SDOF system and the seismic demand (displacement at the top floor of the frame) as recommended by the codes.

For the evaluation of the seismic demand in the following the procedure proposed in the Italian Code [30] is employed. In particular, the spectrum here adopted for the worked examples is computed considering the data associated to the city of Catania (Italy), soil type B, modal damping equal to 5%. The considered spectrum which is associated to a PGA equal to 0,283 g, is reported in the following Figure 4.

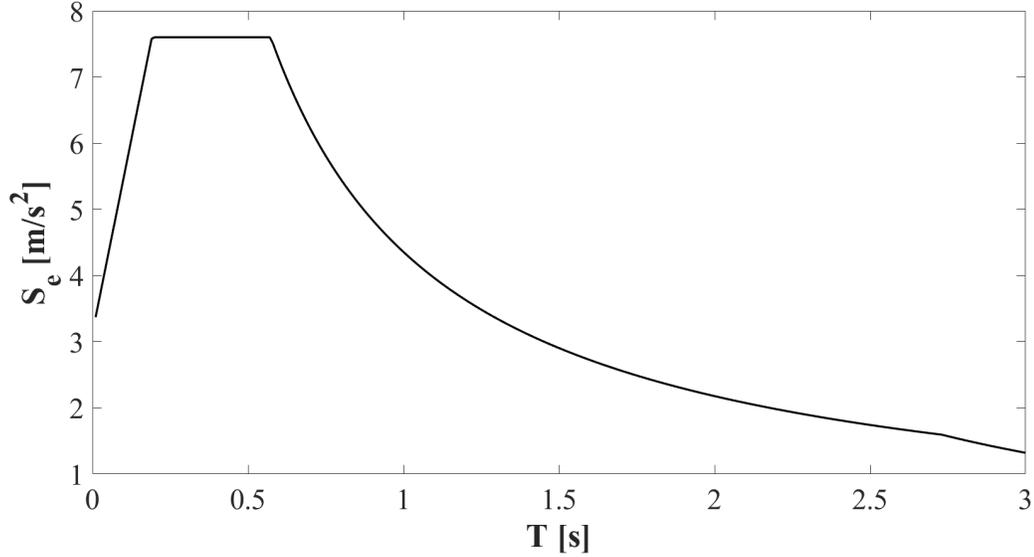

Figure 4. Design spectrum adopted to assess the seismic vulnerability.

For each profile chromosome $C_k^{ext}$ two internal genetic algorithms ("Internal GA A" and "Internal GA B") are firstly launched for the determination of the collapse loads and failure mechanisms when mass proportional (A) or inverse triangular (B) horizontal load distributions are applied. The consequent two values of the collapse loads, say $\lambda_c^A$ and $\lambda_c^B$, are the minimum ones corresponding to the winning chromosomes $C_A^{int}$ and $C_B^{int}$, obtained for the considered load distributions after several runs of the algorithms (typically $N^{int}$=10, in order to avoid local minima). When the two collapse loads are evaluated, the corresponding safety factors $SF_k^A$ and $SF_k^B$ of the profile chromosome $C_k^{ext}$ with respect to expected seismic demand are calculated.

## 2.3 Fitness calculation

In the identification of the most performant frame many requirements must be considered. One of these is of course that both the two safety factors described in the previous paragraph must be greater than one. Other considerations, involving for example the economic cost of the structural members (related to the weight of the frame) and further engineering design precautions, can be introduced in the definition of the following fitness value $F_k$ of the profile chromosome $C_k^{ext}$ :

$$F_k = \alpha_1 f_{1k} + \alpha_2 f_{2k} + \alpha_3 f_{3k} + \alpha_4 f_{4k} \qquad (5)$$

where

$f_{1k} = 1 - 0.5 \left( \frac{1}{SF_k^A} + \frac{1}{SF_k^B} \right)$ is the fitness component related to the seismic safety factors;

$f_{2k} = 1 - \frac{Mass_k}{Max\,Mass}$ takes into account the ratio between the mass ($Mass_k$) of the frame associated to the profile chromosome $C_k^{ext}$ and the maximum one (*Max Mass*) which corresponds to a frame with all the structural members having the biggest cross section in the list of considered profiles;

$f_{3k} = \frac{1}{2N} \left( \sum_{j=1}^{N} c_j^A + \sum_{j=1}^{N} c_j^B \right)_k$ allows to control the failure mode privileging the frames for which more elementary mechanisms (and therefore more unitary values of $c_j$) are involved, that is the failure mode is global, which implies a large energy dissipation;

$f_{4k} = \frac{1}{Nf-1} \left[ \sum_{ic=1}^{Nf-1} \mathcal{F}(c_{ic} - c_{(i+1)c}) \right]_k$ , where $\mathcal{F}(x-y) = 1$ if $x > y$ and $\mathcal{F}(x-y) = -1$ otherwise, is the fitness component that takes into account a resistance hierarchy according to which the cross section of the columns cannot increase with the height of the frame.

The definitions of the four components of the fitness imply that the four values range between 0 and 1, thus favouring the introduction of the relative weights $\alpha_1, \alpha_2, \alpha_3, \alpha_4,$ by means of which the designer can calibrate the above requirements.

Once the fitness of all the profile chromosomes $C_k^{ext}$ ($k = 1, ..., P^{ext}$) have been evaluated by means of a parallel computing procedure, the fitness array is processed by the External GA which produces the next generation. The process is repeated until the established number of iterations $N^{ext}$ has been reached. The optimal design of the frame is finally achieved considering the set of cross sections for beams and columns associated to the highest fitness.

## 3   Algorithm implementation

The algorithm as a whole matches the classic case of a job workflow running $N^{ext} \bullet P^{ext}$ times a direct acyclic graph (DAG) where $N^{ext}$ expresses the number of the External GA generations and $P^{ext}$ the number of profile chromosomes to compute at each cycle. Although several commercial software products are available to implement this kind of computational workflow, their use may be better justified in those cases where the process flow may vary during the computation. In the case of the nested genetic algorithm, from now on briefly called Optimal Design (OD), the computational workflow ever has a static configuration so that it can be easily controlled by a dedicated set of shell scripts. In particular, as shown in Figure 2, a main script is in charge of controlling the NetLogo External GA and a number of worker scripts are in charge to execute the necessary NetLogo Internal GA, A and B, algorithms. The worker script also computes the final fitness $F_k$ value as soon as all $\lambda_c^A$ and $\lambda_c^B$ will be available for each chromosome $C_k^{ext}$. Once all fitness values have been determined, the main script wakes up the External GA and a new generation calculation will start till the maximum number $N^{ext}$ of repetitions is reached. The execution coordination among the main and workers scripts is guaranteed by a database instance which contains just two tables, the first used by workers, holding a task queue and a chromosomes table populated by workers, containing relevant computed values such as the fitness $F_k$.

The computation of the OD makes use of the virtualization based on docker containers. Since the earlier design phases, the virtualization has been considered a key element, because it offers the possibility to run on a wide range of physical platforms, especially those provided by open and private Cloud infrastructures. This way to access computing infrastructures also becomes a key enabler to let this work be reproducible and reusable, two important factors in the incoming age of Open Science ([www.fosteropenscience.eu/node/1420](www.fosteropenscience.eu/node/1420)) and FAIR principles ([www.go-fair.org/fair-principles/](www.go-fair.org/fair-principles/)) compliant research data. For this reason, all source codes related to the running environment setup and the algorithm execution are publicly available in the INFN source code repository, based on GitLab (baltig.infn.it) inside the 'optimal design' project (baltig.infn.it/brunor/optimal_design).

The decision of using docker ([www.docker.com](www.docker.com)) to manage the virtualized resources has been made for principally two reasons. The first is associated with the simplicity to install, use, and maintain its environment. The second is related to its software maturity and popularity. For these reasons, docker can be considered today a standard technology in the world of the virtualization based on containers.

The software development, testing and the final data extraction, have been done executing the OD algorithm on top of the GARR Cloud Infrastructure (www.garr.it/en/garr-en/documents/documenti-tecnici/3474-garr-white-paper-maggio-2017), an IaaS provider that uses the virtualization engine OpenStack (www.openstack.org ). The total amount of resources allocated to the Optimal Design project was:

| | |
|---:|:---|
| 66 | Instances |
| 100 | Virtual CPUs |
| 100 GB | Virtual RAM space |
| 1 TB | Virtual storage space |
| 132 | Upmost number of volumes |

Earlier tests have been made on small virtual machines or small-sized virtual clusters, while for the final data extraction computations four virtual machines made of 24CPU, 24GB of RAM each, have been used. All these virtual machines have been connected to the same storage volume via NFS and internally accessible by the same path name: `/data`.

Using the VM configuration above, it was possible to configure the OD services in the following way:

| Service | Instances | Description |
|---------|-----------|-------------|
| db | 1 | Database storing the task queue and profiles chromosomes |
| dbmgr | 1 | A tiny REST-API service used by workers and main computing scripts to perform I/O operations on top of the database |
| master | 1 | Service controlling the NetLogo External GA |
| worker | 15 | Service in charge to execute NetLogo Internal GA 1 and 2 algorithms and perform the final Fitness calculation. |
| Web | 1 | Service used to expose an execution dashboard, a web-dav access to the computed data, and several other minor services. |

# 4 Applications

In the present paragraph the proposed design strategy is applied to two different steel frames with respectively 2 and 5 storeys. Therefore, the profile chromosome size is 4 and 10, respectively. The numbers of the external genetic algorithm generations and of the chromosomes to compute at each cycle have been chosen respectively equal to $N^{ext}=30$ and $P^{ext}=100$. The correspondent values for the internal (A and B) algorithms are $N^{int}=40$ $P^{int}=100$. Three subsections are here considered; in the first one, the computational effort required by the proposed procedure is discussed; then, the two worked examples are described and analysed in the next two subsections.

## 4.1 Computational analysis of the proposed strategy

In order to provide a computing time estimation for the two considered frames which would not depend on the number of available resources, the results summarized in Table 1 have been extracted for a single profile chromosome, considering the average computing time (expressed in seconds) of the internal algorithms (A and B) and the fitness calculation, extracted from the task queue.

Table 1. Computational effort needed by the proposed procedure.

| Frame | Algorithm | Average seconds |
|---|---|---|
| **2 storey frame** | Internal GA A; Internal GA B (single run) | 10.34; 10.27 |
| | Fitness Calculation | 10.18 |
| | Total time for 1 profile chromosome | 20.48 |
| | **Total time for the whole procedure** | **614.40** |
| **5 storey frame** | Internal GA A; Internal GA B (single run) | 85.35; 85.13 |
| | Fitness Calculation | 10.25 |
| | Total time for 1 profile chromosome | 95.49 |
| | **Total time for the whole procedure** | **2864.70** |

In the table, for each frame, the total average computing time per chromosome is reported as the sum of the two consecutive steps: internal GA time (for a single run, further averaged over A and B) + fitness calculation time. Since each one of the $P^{ext} = 100$ profile chromosomes is elaborated in parallel during a single generation (as well as the 10 runs of each Internal algorithm), the total average computing time for the whole procedure can be obtained simply multiplying the previous total time for the $N^{ext} = 30$ generations of the external GA and is also reported in bold for both the cases.

Of course, these values hold for unlimited (and homogeneous) resources, but in real cases they depend on the number of cores effectively available in the virtual infrastructure and on their performance. Furthermore, it must be pointed out that computing time strongly depends also on the kind of physical resources beyond the virtualized ones, here provided by the GARR institute; this time can be different if other cloud providers are adopted or if physical resources are directly used.

## 4.2 Two-storey frame

The considered frame, reported in Figure 5, has 2 bays each one of length 4 m and 2 storeys with interstorey height equal to 3 m. The beams are subjected to a permanent distributed vertical load equal to 50 kN/m. According to standard seismic analyses two horizontal force distributions are considered, namely a mass proportional force distribution and an inverse triangular one. The relevant horizontal forces are summarized in Table 2.

Table 2. Horizontal force distributions (in kN) for the two-storey frame.

|  | Force distribution | |
|---|---|---|
|  | Mass proportional | Inverse triangular |
| $F_1$ | 400 | 266.67 |
| $F_2$ | 400 | 533.33 |

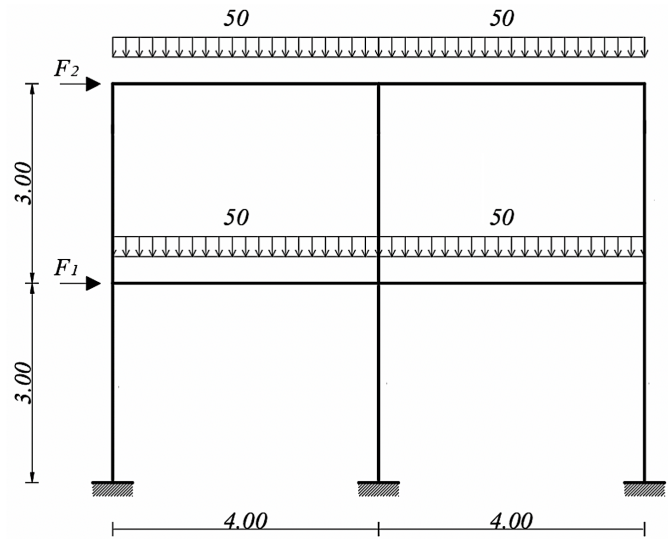

Figure 5. First benchmark planar frame (length in m, forces in kN)

The objective of the design strategy is to select the best set of profiles for beams and columns among a predefined list. In the present application reference is made to steel profiles in the class HEA and 10 possible cross sections are considered as reported in Table 3. Axial and shear deformability are neglected. The Young's modulus adopted for the steel is equal to 210000 MPa.

For the considered frame, the generic profile chromosome for the application of the external genetic algorithm has 4 genes corresponding respectively to the ID of the profile of the beams and columns at the first and second floor. Since each gene can assume ten different values, the overall number of possible different chromosomes in the population is $P_{max} = 10^4$. The adopted relative weights are $\alpha_1=0.2$, $\alpha_2=0.6$, $\alpha_3=0.1$ and $\alpha_4=0.1$.

Table 3. Predefined list of the considered set of adopted cross sections

| ID | Profile | I [cm⁴] | A [cm²] | $W_p$ [cm³] | $M_p$ [KNm] |
|---|---|---|---|---|---|
| 1 | HE 140A | 1033 | 10.2 | 173.5 | 40.7725 |
| 2 | HE 160A | 1673 | 13.21 | 245.1 | 57.5985 |
| 3 | HE 180A | 2510 | 14.47 | 324.9 | 76.3515 |
| 4 | HE 200A | 3692 | 18.08 | 429.5 | 100.9325 |
| 5 | HE 220A | 5410 | 20.67 | 568.5 | 133.5975 |
| 6 | HE 240A | 7763 | 25.18 | 744.6 | 174.981 |
| 7 | HE 260A | 10450 | 28.76 | 919.8 | 216.153 |
| 8 | HE 280A | 13670 | 31.74 | 1112 | 261.32 |
| 9 | HE 300 A | 18260 | 37.28 | 1383 | 325.005 |
| 10 | HE 320A | 22930 | 41.13 | 1628 | 382.58 |

Performing the proposed nested algorithm strategy, several profile chromosomes are analysed and the one with the highest fitness will be considered as the winner. In Table 4 the five profile chromosomes correspondent to the highest values of the fitness are reported together with each rate defined in Eq.(5). Aiming at a deeper insight on the differences between the above results, more details related to the first two profile chromosomes shown in the table are provided in the following. In particular, for each one of the two correspondent frames, the value of the collapse load multiplier $\lambda_c$, the correspondent collapse mechanism, the ultimate displacement of the equivalent elastic-plastic SDOF system $d_{cu}$, the seismic demand $d_{max}$ and the related safety factors SF for the two considered load conditions are reported in Table 5.

Table 4. Five winning profile chromosomes for the two-storey frame

| Profile chromosome | Fitness | $f_1(\alpha_1=0.2)$ | $f_2(\alpha_2=0.6)$ | $f_3(\alpha_3=0.1)$ | $f_4(\alpha_4=0.1)$ |
|:---:|:---:|:---:|:---:|:---:|:---:|
| [5 5 3 4] | 0.613 | 0.641 | 0.542 | 0.6 | 1 |
| [4 4 3 4] | 0.608 | 0.422 | 0.574 | 0.8 | 1 |
| [5 5 4 4] | 0.601 | 0.656 | 0.516 | 0.6 | 1 |
| [6 5 4 4] | 0.594 | 0.601 | 0.491 | 0.8 | 1 |
| [4 5 3 4] | 0.589 | 0.598 | 0.559 | 0.35 | 1 |

In the same table, in the bottom panels, a visual representation of the temporal spreading of the winning profiles among the population of chromosomes is reported for the two cases. In particular, the population of $P^{ext}= 100$ profile chromosomes, indicated by coloured cells (different colours correspond to different chromosomes), is reported in the *x*-axis, while in the *y*-axis (from bottom to top) the sequence of the $N^{ext} = 30$ generations of the external GA is shown: the temporal convergence towards a single colour expresses the diffusion of the winning chromosome among the population. The single coloured cells, randomly distributed in the top part of the figure, represent random mutations of the winning chromosome produced by the GA.

It is interesting to highlight that the fitness of the profile chromosome [5 5 3 4] turns out to be greater than that of [4 4 3 4] because the former has higher values of the safety factors in both the load conditions. Nevertheless, it must be pointed out that the latter profile chromosome is associated to smaller dimensions of the cross sections of columns and beams at the first floor and therefore its realization is economically more convenient. This consideration allows to underline that the provided results are deeply related to the weights introduced for each component of the fitness function which can be conveniently calibrated according to desired requirements.

Table 5. Comparison between two profile chromosomes of the two-storey frame

| | [5 5 3 4] | [4 4 3 4] |
|---|---|---|
| $\lambda_c^A$ | 0.301 | 0.239 |
| *Collapse Mech_A* | 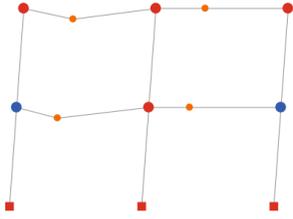 | 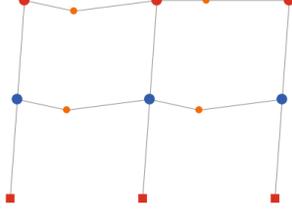 |
| $d_{cu}^A$ | 0.247 | 0.186 |
| $d_{max}^A$ | 0.114 | 0.140 |
| $SF^A$ | 2.167 | 1.324 |
| $\lambda_c^B$ | 0.271 | 0.215 |
| *Collapse Mech_B* | 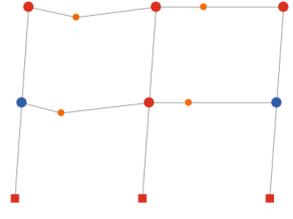 | 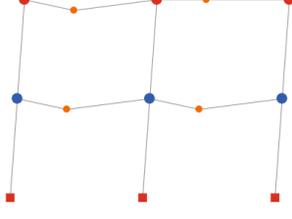 |
| $d_{cu}^B$ | 0.250 | 0.187 |
| $D_{max}^B$ | 0.129 | 0.150 |
| $SF^B$ | 1.945 | 1.250 |
| *Temporal spreading* | 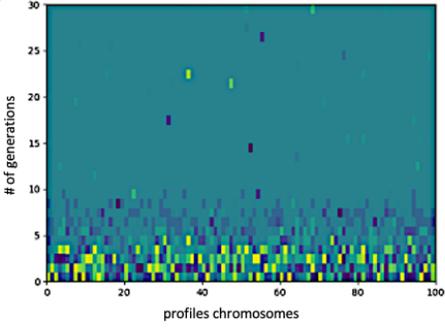 | 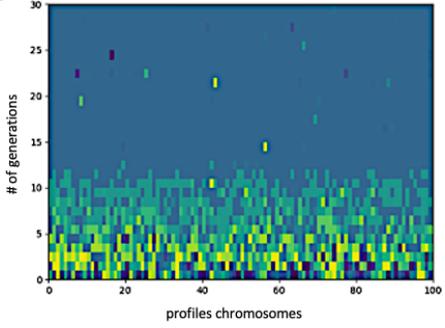 |

## 4.3 Five-storey frame

The second application refers to a frame with three bays and five storeys. The storey height is equal to 3.2 m and the length of the bays is 6.5 m (Figure 6). The beams are subjected to a permanent distributed vertical load equal to 25 kN/m while the horizontal forces in the two considered load conditions are reported in Table 6.

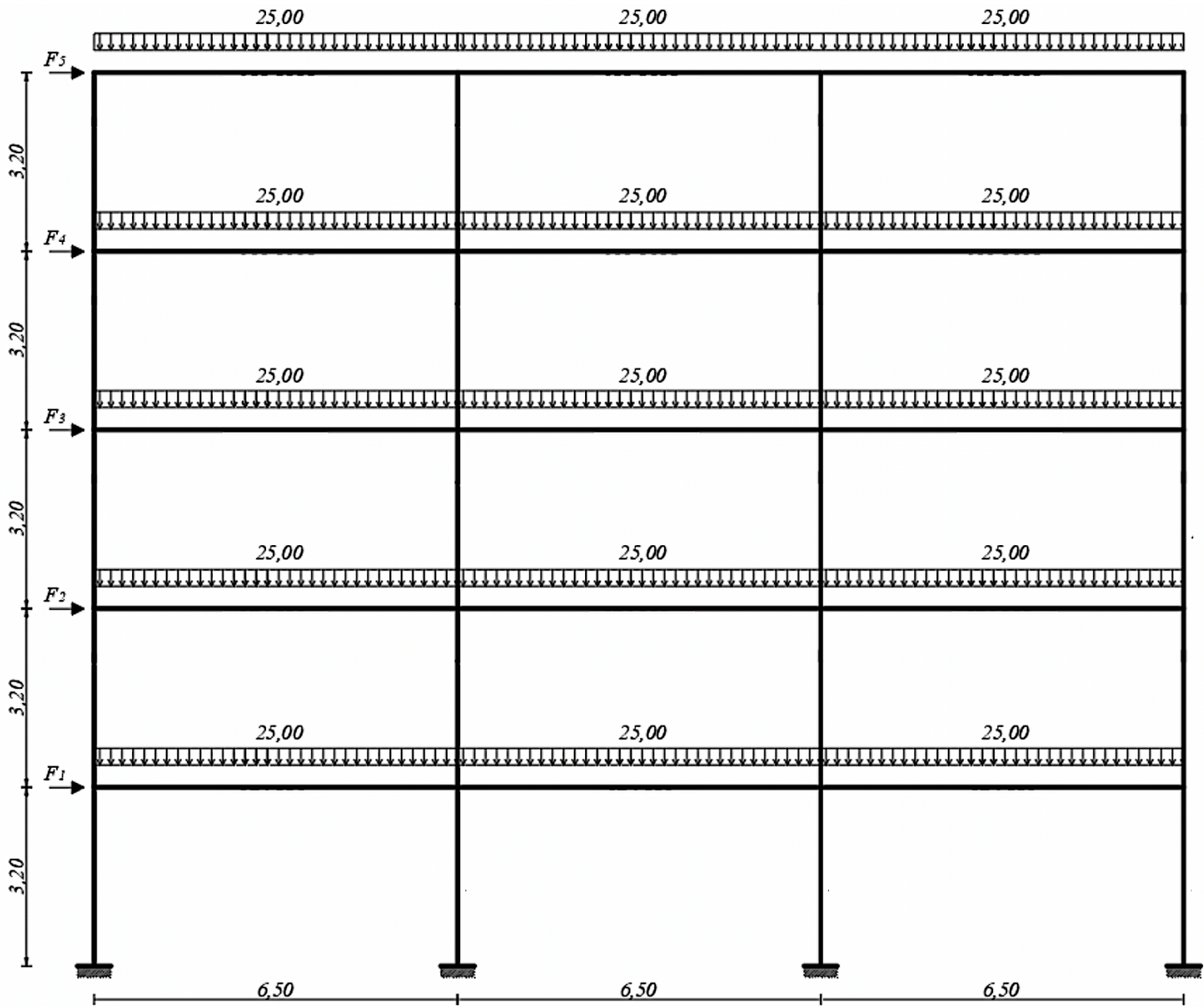

Figure 6. Second benchmark planar frame (length in m, forces in kN)

Table 6. Horizontal force distributions (in kN) for the five-storey frame.

|   | Force distribution | |
|---|---|---|
|   | Mass proportional | Inverse triangular |
| $F_1$ | 325.00 | 108.33 |
| $F_2$ | 325.00 | 216.67 |
| $F_3$ | 325.00 | 325.00 |
| $F_4$ | 325.00 | 433.33 |
| $F_5$ | 325.00 | 541.67 |

Table 7 reports the five profile chromosomes, obtained with the proposed design strategy, which correspond to the highest values of the fitness. Each rate of the fitness defined in Eq.(5) is also reported.

Table 7. Five winning profile chromosomes for the five-storey frame

| Profile chromosome | Fitness | $f_1(\alpha_1=0.2)$ | $f_2(\alpha_2=0.6)$ | $f_3(\alpha_3=0.1)$ | $f_4(\alpha_4=0.1)$ |
|---|---|---|---|---|---|
| [5 3 5 3 5 2 3 2 1 1] | 0.690 | 0.580 | 0.624 | 1 | 1 |
| [6 3 6 3 5 3 4 3 1 2] | 0.687 | 0.738 | 0.582 | 0.895 | 1 |
| [6 3 5 4 5 3 4 3 1 2] | 0.685 | 0.743 | 0.582 | 0.868 | 1 |
| [7 4 5 4 3 3 3 3 1 2] | 0.684 | 0.763 | 0.588 | 0.789 | 1 |
| [8 3 7 3 4 3 4 3 1 2] | 0.677 | 0.744 | 0.563 | 0.894 | 1 |

Again, for this frame a comparison between the two best profile chromosomes is performed and the details are reported in Table 8.

It is interesting to point out that, adopting in the design processes of the two and five storey frames the same values of the relative weights $\alpha_1, \alpha_2, \alpha_3, \alpha_4,$ the safety factors and the total mass play different roles in the identification of the winning profile chromosome. Actually, differently than what happened for the smaller frame described in the previous paragraph, in the case of the five storey frame, the chromosome with the best fitness has smaller safety factors with respect to those of the second one but its total mass (and therefore its economic cost) is smaller.

Table 8. Comparison between two profile chromosomes of the five-storey frame

| | [5 3 5 3 5 2 3 2 1 1] | [6 3 6 3 5 3 4 3 1 2] |
|---|---|---|
| $\lambda_c^A$ | 0.129 | 0.162 |
| *Collapse Mech_A* | 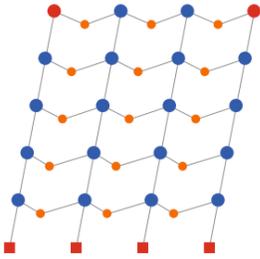 | 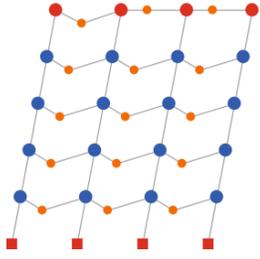 |
| $d_{cu}^A$ | 0.383 | 0.535 |
| $d_{max}^A$ | 0.207 | 0.183 |
| SF$^A$ | 1.850 | 2.930 |
| $\lambda_c^B$ | 0.105 | 0.133 |
| *Collapse Mech_B* | 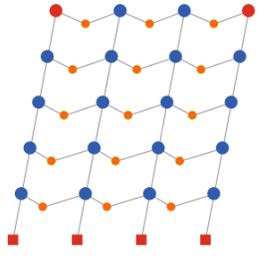 | 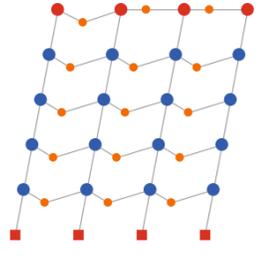 |
| $d_{cu}^B$ | 0.379 | 0.531 |
| $d_{max}^B$ | 0.226 | 0.193 |
| SF$^B$ | 1.675 | 2.747 |
| *Temporal spreading* | 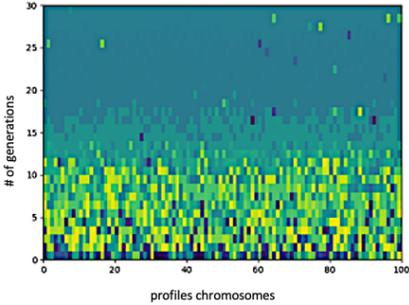 | 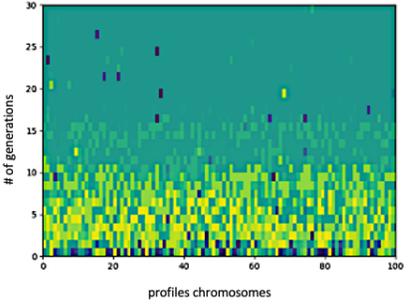 |

In fact, as it can be easily noticed, all the structural members of the winning profile chromosome have cross sections smaller, or at least equal, to those of the frame described by the second best chromosome. It is worth noticing anyway that the difference between the values of the fitness of the two best chromosomes is small and therefore, as already pointed out, a small change in the relative weights could lead to different final results. Therefore, according to his wishes, the designer can privilege one of the introduced requirements ore add some new ones in the definition of the fitness of a profile chromosome.

## 5  Conclusions

The present paper proposes an original multi-objective strategy for the optimal plastic design of frames subjected to seismic excitations. The procedure is based on the application of two different genetic algorithms launched in a nested structure. In particular, when the geometry (lengths of the beams and of the columns) and loads acting on the frame that must be designed are assigned, the external algorithm is able to explore among different configurations associated to different size of the cross sections of the member. Each configuration is associated to a fitness that considers many engineering requirements for an optimal design and embeds the modern concepts to assure a good overall behaviour of the structure (e.g. global ductility, hierarchy criterion, high energy dissipation in seismic conditions). One of these requirements is the safety factor of the considered frame with respect to seismic expected excitations and this is calculated by means of the internal genetic algorithm. In addition, the economic cost in the realization of the frame, related to its total mass, is accounted for. The application of the proposed procedure to two steel frames of different size allows highlighting the importance of the definition of the fitness and of the weights associated to each one of its components. The proposed procedure must be therefore considered as a proposal for a new design strategy which can be either enriched taking into account other fitness components or opportunely calibrated according to specific requirements. It is worth noticing that the software implementation has been conceived in order to let this work be reproducible and reusable, according to the Open Science paradigm and FAIR principles.

## Acknowledgments

This research was funded by the University of Catania, with the projects "Linea di intervento 2 e Starting Grant del Piano di incentivi per la ricerca di Ateneo 2020/2022 " of the Departments of Civil Engineering and Architecture and Physics and Astronomy "Ettore Majorana" and by the Italian Ministry of University and Research with the project "PRIN2017 linea Sud: Stochastic forecasting in complex systems". Part of the resources used in this work have been provided by the Cloud infrastructure at GARR, the Italian Research and Education Network.